\documentclass[conference]{IEEEtran}
\usepackage[cmex10]{amsmath}
\usepackage{times,float, cite,subfigure,graphicx,epsfig,wrapfig,amsfonts,amssymb,algorithmic,algorithm,threeparttable,color,amsthm}
\usepackage[dvips,bookmarks=false,citecolor=blue,urlcolor=blue]{hyperref} 

\usepackage[T1]{fontenc}
\usepackage[USenglish,UKenglish,french,spanish,italian]{babel}
\usepackage[nodayofweek,level]{datetime}

\newcommand{\mydate}{\formatdate{27}{7}{2015}}

\begin{document}

%

\begin{titlepage}

\begin{tabular}{l        r}

\includegraphics[bb=10bp 0bp 600bp 750bp,clip,scale=0.3]{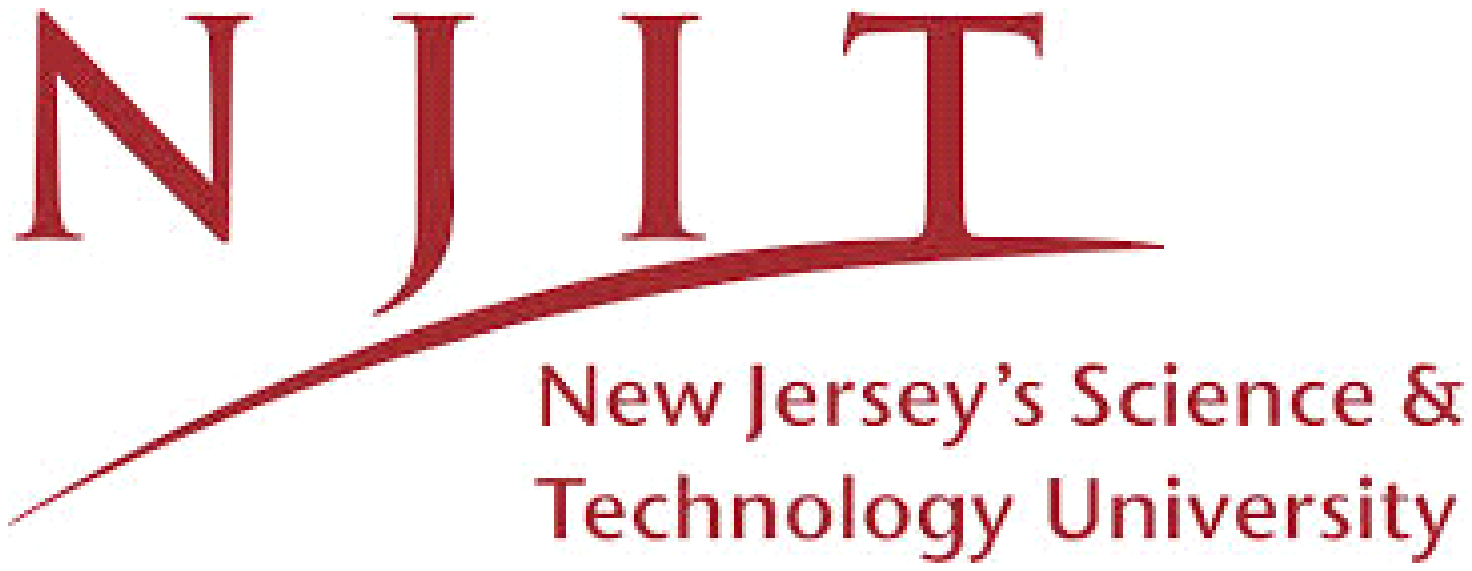} \hspace{6cm} & \includegraphics[bb=0bp -200bp 500bp 950bp,clip,scale=0.2]{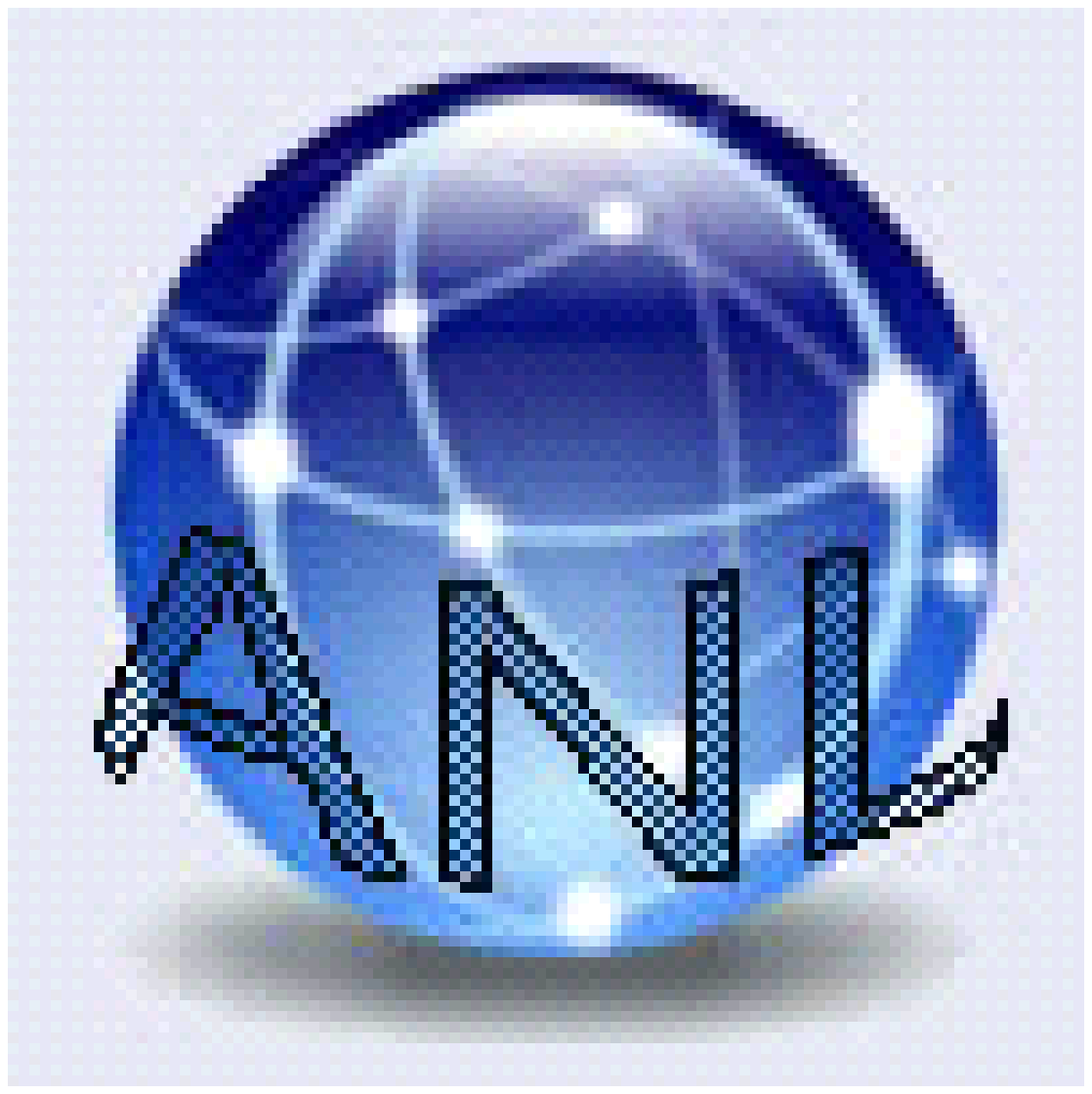}

\end{tabular}

\begin{center}

\textsc{\LARGE Design and Analysis of Green Optical Line Terminal for TDM Passive Optical Networks}\\[1.5cm]

{\Large \textsc{mina taheri}}\\ 
{\Large \textsc{nirwan ansari}}\\[2cm] 

{}
{\textsc{TR-ANL-2015-004}\\
\selectlanguage{USenglish}
\large \mydate} \\[3cm]

{\textsc{Advanced Networking Laboratory}}\\
{\textsc{Department of Electrical and Computer Engineering}}\\
{\textsc{New Jersy Institute of Technology}}\\[1.5cm]
\vfill

\end{center}

\end{titlepage}


\begin{abstract}
This paper proposes a novel scheme which can efficiently reduce the energy consumption of Optical Line Terminals (OLTs) in Time Division Multiplexing (TDM) Passive Optical Networks (PONs) such as EPON and GPON. Currently, OLTs consume a significant amount of energy in PON, which is one of the major FTTx technologies. To be environmentally friendly, it is desirable to reduce energy consumption of OLT as much as possible; such requirement becomes even more urgent as OLT keeps increasing its provisioning data rate, and higher data rate provisioning usually implies higher energy consumption. In this paper, we propose a novel energy-efficient OLT structure which guarantees services of end users with the smallest number of power-on OLT line cards. More specifically, we adapt the number of power-on OLT line cards to the real-time incoming traffic. Also, in order to avoid service disruption resulted by powering off OLT line cards, proper optical switches are equipped in OLT to dynamically configure the communications between OLT line cards and ONUs.
\end{abstract}


\section{Introduction}
As energy consumption is becoming an environmental and therefore social and economic issue, green Information and Communication Technology (ICT) has attracted significant research attention recently. It was reported that Internet consumes as much as $\sim 1\%-2.5\% $ of the total electricity in broadband enabled countries\cite{BalEne09, pickavet2009worldwide, fettweisict}, and currently and in the medium term future, the majority of the energy of Internet is consumed by access networks owing to the large quantity of access nodes \cite{BalEne08}.

Energy consumptions of access networks depend on the access technologies. Among various access technologies including WiMAX, FTTN, and point to point optical access networks, passive optical networks (PONs) consume the smallest energy per transmission bit attributed to the proximity of optical fibers to the end users and the passive nature of the remote node \cite{LanOnt08}. However, as PON is deployed worldwide, it still consumes a significant amount of energy. It is desirable to reduce the energy consumption of PONs since every single watt saving will end up with overall terawatt and even larger power saving. Reducing energy consumption of PONs becomes even more important as the current PON systems evolve into next-generation PONs with increased data rate provisioning \cite{ZhaNex09, ansari2013media}.

In PONs, energy is consumed by optical line terminal (OLT) and optical network units (ONUs). Owing the large quantity, ONUs consume a large portion of the overall PON energy \cite{ZhaTow11}. Although OLT consumes a less amount of power than the total aggregated ONUs, one OLT line card does consume a much larger amount of power than one ONU. Reducing energy consumption of OLT is as important as reducing energy consumption of ONUs especially from the operators' and home users' perspectives. For the network operators, decreasing the energy consumption of OLT can significantly reduce the energy consumption of the central office, while decreasing the energy consumption of ONUs has small and likely negligible impacts on that of home users who have many other electrical appliances with much higher energy consumption.

Formerly, sleep mode and adaptive line rate have been proposed to efficiently reduce the power consumption of ONUs by taking advantages of the bursty nature of the traffic at the user side \cite{zhang2013standards, KubStu10, WonSle09, ChoEne10, taheri2014multi}. It is, however, challenging to introduce ``sleep'' mode into OLT to reduce its energy consumption for the following reasons. In PONs, OLT serves as the central access node which controls the network resource access of ONUs. Putting OLT into sleep can easily result in service disruption of ONUs in communicating with the OLT. Thus, a proper scheme is needed to reduce the energy consumption of OLT without degrading services of end users.

In this paper, we propose a novel energy-efficient OLT structure which can adapt its power-on OLT line cards according to the real-time arrival traffic. To avoid service degradation during the process of powering on/off OLT line cards, proper devices are added into the legacy OLT chassis to facilitate all ONUs communicate with power-on line cards. To the best of our knowledge, this is the first work focusing on reducing the energy consumption of OLT \footnote{The preliminary idea was first presented at Sarnoff2011 \cite{Zhang2011}.}.

\section{Framework of the energy-efficient OLT design}
\label{sec:I}

\begin{figure*}[ht!]
\centering
 \includegraphics[scale=0.6]{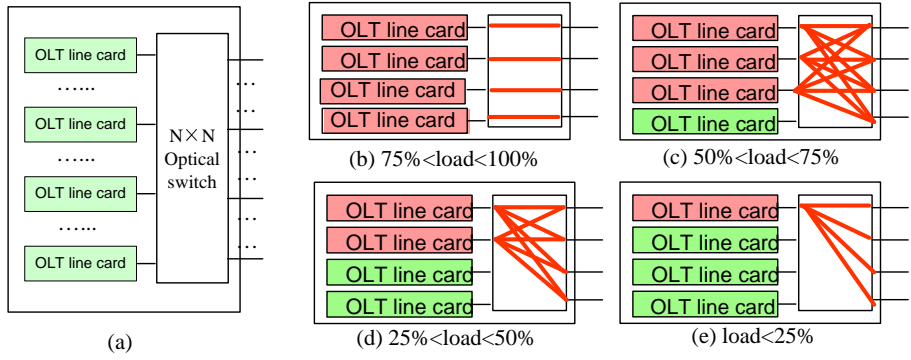}
\caption{OLT with optical switches}
\label{fig:engolt}
\end{figure*}

In the central office, one OLT chassis typically comprises of multiple OLT line cards that transmit downstream signals and receive upstream signals at different wavelengths. Each line card communicates with a number of ONUs. Two wavelengths for the uplink and the downlink are assigned to each ONU. In the currently deployed EPON and GPON systems, one OLT line card usually communicates with either $16$ or $32$ ONUs and such an arrangement is referred to as a PON segment. To avoid service disruptions of ONUs connected to the central office, all these OLT line cards in the OLT chassis are usually power-on all the time. To reduce the energy consumption of OLT, our main idea is to adapt the number of power-on OLT line cards in the OLT chassis to the real-time incoming traffic.

There are two types of subscribers that each network serves; Business subscribers and residential subscribers. Business and residential areas are usually disjunct. It is more likely that each PON segment serves either business customers or residential customers. These two types of customers have different traffic profiles. Business users demand high bandwidth during the day and low bandwidth at night while residential customers request high bandwidth in the evening and low bandwidth during the day.

During the day time, residential segments are lightly loaded. Therefore, one OLT line card can serve several residential segments. In the similar way, the traffic from the business segments can combined to traverse a smaller number of line cards in the evening. 


Business and residential segments usually have low bandwidth demands during the midnight. In these situations, the whole network is lightly loaded. In order to save energy, the number of line cards can be reduced based on the traffic volume. 

Parameters of the proposed model are notated below:
 
$C_u$: Data rate of one OLT line card in the upstream direction.\

$C_d$: Data rate of one OLT line card in the downstream direction.\

$L$: Total number of line cards (PON segments).\

$N_j$: Number of ONUs connected to PON segment $j$.\

$T$: Fixed traffic cycle in TDM PON.

$u_{i,j}(t)$: Arrival upstream traffic rate from ONU $i$ of PON segment $j$ at time $t$.\

$d_{i,j}(t)$: Arrival downstream traffic rate to ONU $i$ of PON segment $j$ at time $t$.\

$l(t)$: smallest number of required OLT line cards at time $t$.\
 
By powering on all the OLT line cards, the overall upstream data rate and downstream data rate accommodated by the OLT chassis equal to $C_u \cdot L$ and $C_d \cdot L$, respectively. $C_u \cdot L$ (or $C_d \cdot L$) may be greater than the real-time upstream (or downstream) traffic.

The traffic rate of each segment cannot be more than the provisioned capacity of the dedicated fiber. Therefore, the following constraints have to be satisfied for any segment $j$:
$$\sum_{i=1}^{N_j}{u_{i,j}(t)} \leq C_u$$ 
$$\sum_{i=1}^{N_j}{d_{i,j}(t)} \leq C_d$$  
The real time incoming upstream and downstream traffics are defined as  $\mathop{\sum_{j=1}^L\sum_{i=1}^{N_j}}{u_{i,j}(t)}$ and $\mathop{\sum_{j=1}^L\sum_{i=1}^{N_j}}{d_{i,j}(t)}$, respectively. 
Then, $$ l(t)= max ({\lceil\mathop{\sum_{j=1}^L\sum_{i=1}^{N_j}}{u_{i,j}(t)}/{C_u}\rceil , \lceil\mathop{\sum_{j=1}^L\sum_{i=1}^{N_j}}{d_{i,j}(t)}/C_d\rceil })$$



Our ultimate objective is to \textbf{power on only $l(t)$ OLT line cards to serve all $N$ ONUs} at a given time $t$ instead of powering on all $L$ line cards. However, powering off OLT line cards may result in service disruptions of ONUs communicating with these OLT line cards. To avoid service disruption, power-on OLT line cards should be able to provision bandwidth to all ONUs connected to the OLT chassis. To address this issue, we propose several modifications over the legacy OLT chassis to realize the dynamic configuration of OLT as will be presented next.

\section{OLT with optical switch}
To dynamically configure the communications between OLT line cards and ONUs, one scheme we propose is to place an optical switch in front of all OLT line cards as shown in Fig.\ref{fig:engolt} (a). The function of the optical switch is to dynamically configure the connections between OLT line cards and ONUs. When the network is heavily loaded, the switches can be configured such that each PON system communicates with one OLT line card. 

As discussed in Section \ref{sec:I}, when the network is lightly loaded, the switches can be configured such that multiple PON systems communicate with one line card. Then, some OLT line cards can be powered off, thus reducing energy consumption.

Assume the energy consumption of the optical switch is negligible. As compared to the scheme of always powering on all $L$ line cards, the scheme of powering on only 
$l(t)$ line cards at time $t$ can achieve relative energy saving as large as
$$1-\frac{l(t)}{L}$$

Then, the average energy saving over time span $T$ equals to 
$$\frac{1}{T}{\int_{t=0}^{T}\frac{1-l(t)}{L}dt}$$

We define the traffic load as the maximum of upstream and downstream traffic loads to determine the required number of line cards: $$load = max (\mathop{\sum_{j=1}^L\sum_{i=1}^{N_j}}{u_{i,j}(t)}/{C_u} L \;  , \; \mathop{\sum_{j=1}^L\sum_{i=1}^{N_j}}{d_{i,j}(t)}/{C_d L} )$$
The process of changing the switch configuration is time consuming, frequent change of each line card's status may degrade the ONU performance. Since the traffic of the ONUs is also  bursty and changes dynamically, it is more efficient to monitor the traffic for an observation period ($T_O$) before changing the switch configuration. 

If the traffic load of the network remains below a threshold for $T_O$, the number of line cards will  be adjusted accordingly. By dynamically configuring switches, the number of power-on OLT line cards is reduced from $L$ to $x$ when the traffic load falls between $(x-1)/L$ and $x/L$ for a period of $T_O$. Thus, a significant amount of power can be potentially saved.

Fig. \ref{fig:engolt} (b)-(e) illustrates the configuration of switches for the case that one OLT chassis contains four OLT line cards. The number of power-on OLT line cards is reduced from four to three, two, and one when the traffic load falls within the range $[50\%,75\%]$, $[25\%,50\%]$, and  $[0,25\%]$, respectively.

By equipping the OLT chassis with proper optical switches, the communications between OLT line cards and ONUs can be dynamically configured. The new OLT structure appears to be promising and cost-effective as compared to the WDM based solutions. 
\section{Optical switch specifications}
Configuration time of the optical switches and power consumption of each switch are two main specifications that should be considered in the analysis. 
\subsection{Switch configuration time}
Switches take time to change configurations. The switch reconfiguration time may affect the ONU performances when powering on/off OLT line cards. We investigate the impacts of the switch reconfiguration time on EPON and GPON, respectively.
\begin{itemize}
\item EPON: We argue that services of EPON ONUs are not affected when the switch configuration time is as large as $50$ ms.

In EPON, the upstream bandwidth allocation is controlled by OLT. ONUs transmit the upstream traffic using the allocated time durations stated in the GATE message. IEEE 802.3ah \cite{8023av} specifies that ONUs need to send GATE messages every $50$ ms to maintain registration even if they do not have traffic to transmit. Thus, the disrupted servicing time can be transparent to EPON ONUs when the switch reconfiguration time is as large as $50$ ms, which can be satisfied by most optical switches.
\item GPON: We argue that services of GPON ONUs are not affected when the switching time is no greater than $125 \mu s$.

For GPON, ITU-T G. 984.3 \cite{984} specifies a fixed frame length of 125 $\mu s$. An ONU receives downstream control messages and sends its upstream data or control traffic every GPON frame, i.e., 125 $\mu s$. Therefore, services of GPON ONUs are not affected when the switch configuration time is no greater than $125 \mu s$.
\end{itemize}

\subsection{Power consumption of optical switches}

So far, we have not considered the impact of the power consumption of optical switches on the saved energy of OLT chassis. Optical switches consume some power, and the nonzero power consumption of optical switches may reduce the saved energy of the OLT chassis.

Denote $p(s)$ and $p(l)$ as the power consumption of optical switch and one OLT line card, respectively. By incorporating considering the energy consumption of the optical switch, the power consumption of the proposed OLT chassis at any given time $t$ equals to 
$$p(l)\cdot l(t) + p(s)$$ Hence, the energy saving is reduced to 
$$1-\frac{p(l)/T\int_{t=0}^T{l(t)}dt+p(s)}{p(l) \cdot L}$$ The proposed OLT chassis can save some energy as compared to the legacy OLT chassis when the power consumption $p(s)$ of the optical switch satisfy the following condition 
$$p(s)<(L-\frac{1}{T}\int_{t=0}^T{l(t)}dt)\cdot p(l)$$ The power consumption of optical switch differs from the switch size as well as the manufacturing technology. It is reported that one $2\times 2$ optomechanic switch consumes around $0.2$ w \cite{opto}, while an OLT line card consumes around $5$ w \cite{Vereecken2011}. Then, for the $L=2$ case, energy can be saved as long as the average number of power-on OLT line cards 
$\frac{1}{T}\int_{t=0}^T{l(t)}dt$ is less than $1.96$. This condition can be easily satisfied. Therefore, even considering the power consumption of optical switches, the proposed OLT structure can still achieve a significant amount of power saving as compared to the legacy OLT.

\section{System model}

\begin{figure*}[!ht]
\centering
 \includegraphics[bb=40bp 100bp 680bp 430bp,clip,scale=0.5]{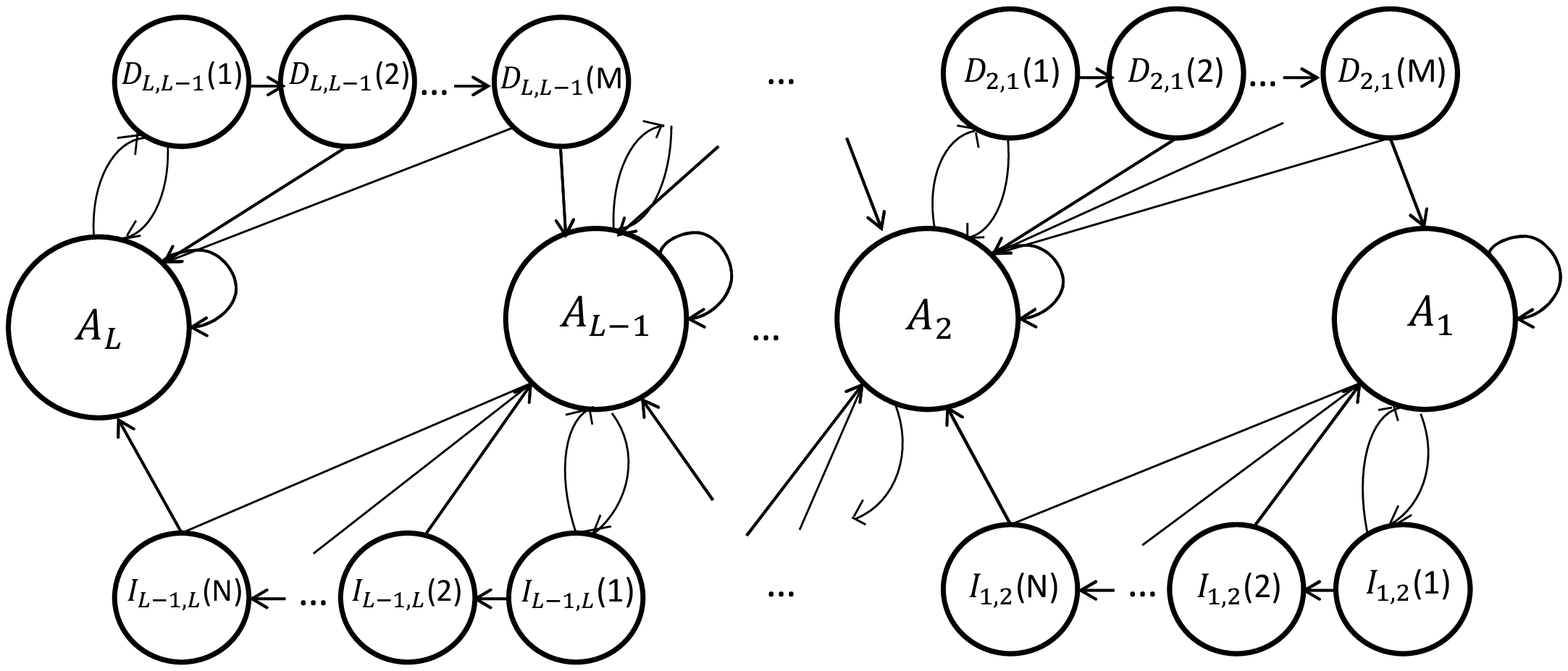}
\caption{State transition in Markov chain.}
\label{fig:mrkv}
\end{figure*}
We employ semi-Markov chains to analyze the system. Different observation periods have been considered for decreasing and increasing the number of line cards. $T_D$ is the observation time period for decreasing the number of active line cards. $T_I$ is defined as an observation period for increasing the number of line cards. In order to ease the analysis, from now on, the formulations are based on the downstream traffic only. Poisson processes with the average rate of  $\lambda_a$ packets per second is assumed for downlink packet arrivals. A fixed traffic scheduling cycle is assumed in the performance analysis. Consider $P_a(\alpha,T)$ as the probability that $\alpha$ downstream packets arrive at the OLT chassis 
during $T$ traffic scheduling cycle:
$$P_a{(\alpha,T)}=e^{-\lambda_a T} \cdot \frac{(\lambda_a T)^ \alpha}{\alpha!}$$ 

The same process can be considered for uplink packet arrival in the case of having downstream and upstream traffic simultaneously. 
 
\subsection{OLT chassis state}

In the defined Markov chain model, the total number of active line cards vary from $1$ to $L$. Therefore, there are $L$ possible states of active line cards. $A_i$ represents $i$ active line cards. Transition from each active state to another active state should be done through the listening states. There is no direct transition between active states. There are two different types of listening states; $D$ (listening states for decreasing the number of line cards) and $I$ (listening states for increasing the number of line cards). 
Figure \ref{fig:mrkv} shows the state transitions.

$D_{i,i-1}(j)$ refers to the state that the total traffic load of the OLT chassis remains below $i/L$ for $j$ cycles. The number of active line cards during $M=T_D/T$ time cycles remains as $i$ line cards. If the traffic load goes higher than  $i/L$ amount during a traffic scheduling cycle, the transition to $A_i$ occurs. Otherwise, the transition from $D_{i,i-1}(j)$ to $D_{i,i-1}(j+1)$ happens. Whenever the number of listening cycles reaches to $M$, the OLT chassis switches to the next lower active state ($A_{i-1}$).

$I_{i, i+1}(k)$ refers to the state that the total traffic load of the OLT chassis remains above $i/L$ for  $k$ time cycles. The number of active line cards during $N=T_I/T$ remains as $i$ line cards. The excess amount of the traffic will be buffered during $T_I$. If the traffic load becomes lower than  $i/L$ amount during a traffic scheduling cycle, the transition to $A_i$ occurs. Otherwise, the transition from $I_{i,i+1}(k)$ to $I_{i, i+1}(k+1)$ happens. If the total traffic load stays beyond the defined threshold ($i/L$) for $N$ traffic scheduling cycles, the OLT chassis switches to the next upper active state ($A_{i+1}$).

\subsection{State transitions}
\begin{itemize}

\item State transitions from $A_i$ to $D_{(i,i-1)}(1)$, $D_{(i,i-1)}(j)$ to $D_{(i,i-1)}(j+1)$, or from $D_{(i,i-1)}(M)$ to $A_{i-1}$  $for \medspace i=2,..,L$ $and$ $j=1,...,M-1$ happen 
when the traffic load in a traffic scheduling time $T$ is less than $(i-1)/L$. Therefore, the probability that the number of arrival packets is smaller than $max=\lfloor (i-1)C_d/Packet \medspace Size \rfloor$ equals to: $\sum_{\alpha=0}^{max}P_a(\alpha,T)$.


\item State transitions from $I_{i,i+1}(k)$ to $A_i$ $for$ $i=2,..,L$ $and$ $k=1,...,N$ happen when the number of arrival packets in a traffic scheduling time $T$ is smaller than $max=\lfloor iC_d/Packet \medspace Size \rfloor$, which is equals to: $\sum_{\alpha=0}^{max}P_a(\alpha,T)$

\item State transitions from $A_i$ to $I_{(i,i+1)}(1)$, from $I_{(i,i+1)}(k)$ to $I_{(i,i+1)}(k+1)$, or from $I_{(i,i+1)}(N)$ to $A_{i+1}$ $for \medspace  i=2,..,L$ $and$ $k=1,...,N-1$ happen when the number of arrival packets in a traffic scheduling time $T$ is greater than $min=\lceil iC_d/Packet \medspace Size \rceil$. The probability equals to $\sum_{\alpha=min}^{\infty}P_a(\alpha,T)$.

\item State transitions from $D_{i,i-1}(j)$ to $A_i$ $for$ $i=2,..,L$ $and$ $j=1,...,M$ happen when the number of arrival packets in a traffic scheduling time $T$ is greater than $min=\lceil (i-1)C_d/Packet \medspace Size \rceil$, which is equals to: $\sum_{\alpha=min}^{\infty}P_a(\alpha,T)$


\item State transition from $A_i$ to itself happens when the number of arrival packets in a traffic scheduling time $T$ remains between $min=\lfloor (i-1)C_d/Packet \medspace Size \rfloor$ and $max=\lceil iC_d/Packet \medspace Size \rceil$. The probability is equal to $\sum_{\alpha=min}^{max}P_a(\alpha,T)$.
\end{itemize}

\subsection{Steady state probabilities}
Denote $P(A_i)$, $P(D_{i,i-1}(j))$ $(i=2,...,L$ $j=1,...,M)$, and $P(I_{i,i+1}(k))$ $(i=1,...,L-1$; $k=1,...,N)$ as the probability of the OLT state when the network is at its steady state. Therefore, the following constraints are satisfied. 	

Steady state probabilities of all the active states except $A(1)$ and $A(L)$ are as follows:
\begin{align} \label{eq:first}
&P(A_i) [pr\{A_i \negmedspace \rightarrow \negmedspace D_{i,i-1}(1)\}+pr\{A_i \negmedspace \rightarrow \negmedspace I_{i,i+1}(1)\}] \nonumber\\
&=[\sum_{k=1}^{\infty}P(I_{i,i+1}(k))pr\{I_{i,i+1}(k) \rightarrow A_i\}]\nonumber\\
&+P(I_{i-1,i}(N))pr\{I_{i-1,i}(N) \rightarrow A_i\}\nonumber\\
&+\sum_{j=1}^{M}P(D_{i,i-1}(j))pr\{D_{i,i-1}(j) \rightarrow A_i\} \nonumber\\
&+P(D-{i+1,i}(M))pr\{D_{i+1,i}(M) \rightarrow A_i\} \nonumber\\
&(i=2,...,L-1)
\end{align}

 Steady state probability of $A_1$ is calculated as follows:
\begin{align}
&P(A_1)pr\{A_1 \rightarrow I_{1,2}(1)\} \nonumber\\
&=\sum_{k=1}^{N}P(I_{1,2}(K)pr\{I_{1,2}(k) \negmedspace \rightarrow \negmedspace A_1\} \nonumber\\
&+P(D_{2,1}(M))pr\{D_{2,1}(M) \negmedspace \rightarrow \negmedspace A_1\}
\end{align}
Steady state probability of $A_L$ is acheived as follows:
\begin{align}
&P(A_L)pr{A_L \rightarrow D_{L,L-1}(1)} \nonumber\\
&=\sum_{j=1}^{M}P(D_{L,L-1}(j))pr\{D_{L,L-1}(j) \rightarrow A_L\} \nonumber\\
&+P(I_{L-1,L}(N))pr\{P(I_{L-1,L}(N)) \rightarrow A_L\}
\end{align}

Steady state probability of listening states $(D)$ except the first and last states are as follows:
\begin{align}
&P(D_{i,i-1}(j))[pr\{D_{i,i-1}(j) \rightarrow D_{i,i-1}(j+1)\} \nonumber\\
&+ pr\{D_{i,i-1}(j)  \rightarrow A_i\}]\nonumber\\
&= \negmedspace P(D_{i,i-1}(j-1))pr\{ \negmedspace D_{i,i-1}(j-1) \negmedspace \rightarrow \negmedspace D_{i,i-1}(j) \negmedspace \} \nonumber\\
&(i=1,...,L) \& (j=2,...,M-1)
\end{align}
Steady state probability of $D_{i,i-1}(1)$ equals to:
\begin{align}
&P(D_{i,i-1}(1))[pr\{D_{i,i-1}(1) \rightarrow D_{i,i-1}(2)\} \nonumber\\ 
&+pr\{D_{i,i-1}(1) \rightarrow A_i\}] \nonumber\\
&=P(A_i)pr{A_i \rightarrow D_{i,i-1}(1)} (i=1,...,L)
\end{align}
Steady state probability of $D_{i,i-1}(M)$ is as follows:
\begin{align}
&D_{i,i-1}(M)[pr\{D_{i,i-1}(M) \rightarrow A_i\} \nonumber\\
&+pr\{D_{i,i-1}(M) \rightarrow A_i-1\}] \nonumber\\
&= \negmedspace P(D_{i,i-1}(M \negmedspace - \negmedspace 1))pr\{ \negmedspace D_{i,i-1}(M \negmedspace - \negmedspace 1) \negmedspace \negmedspace \rightarrow \negmedspace D_{i,i-1}(M)\negmedspace \}
\end{align}
The following is the steady state probability of the listening states $I$ except the first and last states:
\begin{align}
&p(I_{i,i+1}(k))[pr\{I_{i,i+1}(k) \rightarrow A_i\}+ \nonumber\\
&pr\{I_{i,i+1}(k) \rightarrow I_{i,i+1}(k+1)\}] \nonumber\\
&=P(I_{i,i+1}(k-1))pr\{I_{i,i+1}(k-1) \rightarrow I_{i,i+1}(k)\} \nonumber\\
&(i=1,...,L) \& (k=2,...N-1)
\end{align}
Steady state probability of $I_{i,i+1}(1)$ is as follows:
\begin{align}
&P(I_{i,i+1}(1))[pr\{I_{i,i+1}(1) \rightarrow A_i\} \nonumber\\
&+pr\{I_{i,i+1}(1) \rightarrow I_{i,i+1}(2)\}] \nonumber\\
&=P(A_i)pr\{A_i \rightarrow I{i,i+1}(1)\} (i=1,...,L)
\end{align}
Steady state probability of $I_{i,i+1}(N)$ is obtained as follows:
\begin{align} 
&P(I_{i,i+1}(L))[pr\{I_{i,i+1}(L) \rightarrow A_i\} \nonumber\\ 
&+pr\{I_{i,i+1}(L) \rightarrow A_{i+1}\}] \nonumber\\
&=P(I_{i,i+1}(L-1))pr\{I_{i,i+1}(L-1) \rightarrow I{i,i+1}(L)\}
\end{align}

Moreover, the sum of the probabilities of all the states is equal to $1$, i.e.:
\begin{align} \label{eq:last}
\sum_{i=1}^{L}P(A_i) + \sum_{i=1}^{L} \sum_{j=1}^{M} P(D_{i,i-1}(j)) +  \sum_{i=1}^{L} \sum_{k=1}^{N} P(I_{i,i+1}(k))=1
\end{align}

Therefore, the steady state probabilities of all the state can be calculated by solving the above equations \ref{eq:first}-\ref{eq:last}. 

\subsection{Performance analysis}
Denote $p(l)$ as the power consumption of one OLT line card and $P(A_i)$, $P(D_{i,i-1}(j))$ $(i=2,...,L$ $j=1,...,M)$, and $P(I_{i,i+1}(k))$ $(i=1,...,L-1$; $k=1,...,N)$ as the probability of the OLT state when the network is at its steady state. Therefore, the average power consumption equals to:
$$p(l)[\sum_{i=1}^{L}ip(A_i)+\sum_{i=2}^{L}\sum_{j=1}^{M}ip(D_{i,i-1}(j))+\sum_{i=1}^{L-1}\sum_{k=1}^{N}ip(I_{i,i+1}(k))]$$
\begin{figure*}
\centering
 \includegraphics[scale=0.6]{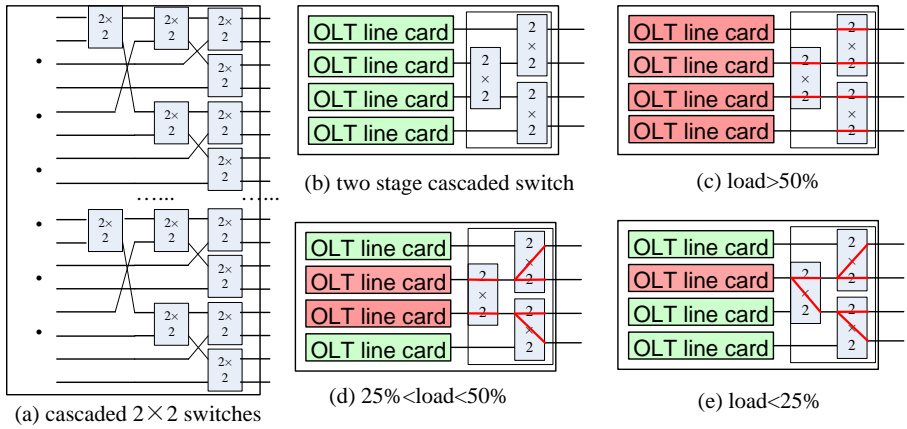}
\caption{OLT with multi-stage cascaded switches}
\label{fig:casc}
\end{figure*}
\section{Cost reduction}
\subsection{OLT with cascaded $2 \times 2$ switches}

Another problem with optical switches is their high prices. The prices of optical switches vary from their manufacturing techniques \cite{MaxOpt03ofc}. At present, there are generally four kinds of optical switches: opto-mechanical switches, micro-electro-mechanical system (MEMS), electro-optic switches, and semiconductor optical amplifier switches. Currently, the opto-mechanic switches are less expensive than the other three kinds. Simply because of their low prices, opto-mechanic switches are generally the adopted choices in designing energy-efficient OLT.

For opto-mechanic switches, an important constraint is their limited port counts. Popular sizes of opto-mechanic switches are $1\times 2$ and $2\times 2$. Considering the port count constraints, we further propose the cascaded $2 \times 2$ switches structure to achieve the dynamic configuration of OLT. More specifically, to replace an $N \times N$ switch, the cascaded $2 \times 2$ switch contains $\log_2^N$ stages and $(N-1)$ $2 \times 2$ switches. Fig. \ref{fig:casc} (a) illustrates the proposed cascaded switches. In the switch, the $k$th stage contains $2^{(k-1)}$ switches.

Fig. \ref{fig:casc} (b) shows a two-stage cascaded $2 \times 2$ switches to replace a $4 \times 4$ switch. As illustrated in Fig. \ref{fig:casc}(c)-(e), when the traffic load is greater than $50\%$, one PON system is connected with one OLT line card; when the traffic load is between $25\%$ and $50\%$, two PON systems are connected with one OLT line card; when the traffic load is less than $25\%$, all PON sytems are connected with a single OLT line card.

Here, we analyze the saved energy of the proposed OLT equipped with cascaded switches. Assume the traffic is uniform among all ONUs. Then, when the traffic load is between $50\%$ and $100\%$, all OLT line cards need be power-on; when the traffic load is between $25\%$ and $50\%$, half of the OLT line cards are powered on. Generally, when the traffic load is between $1/2^k$ and $1/2^{k+1}$, $1/2^k$ of the OLT line cards are power-on. Therefore,
$$1/2^{k+1} \leq load \leq 1/2^k$$
$$k={\lfloor \log_2{(1/load)} \rfloor}$$ 
$$l(t)=1/2^{\lfloor \log_2{(1/load)}\rfloor}$$ The saved energy equals to 
$$1-1/2^{\lfloor \log_2{(1/load)}\rfloor}$$ 
As compared to the OLT with an $N\times N$ switch, the OLT with cascaded $2\times 2$ switches saves a less amount of energy.

Note that the typical switching speed of the opto-mechanical switch is around $5$ ms. As discussed before, it does not affect the performances of users in EPON, but may have impacts on the performances of users in GPON. 

\section{OLT with electrical switches}
Another scheme of avoiding the significant cost increase is to use electrical switches instead. However, before aggregating traffic using electrical switches, the optical transceivers are required to convert the optical signals into electrical signals. Thus, only the energy consumption of the electrical part in an OLT line card can be saved. The energy saving is limited as compared to the scheme of using optical switches. Let $p(e)$ be the energy consumption of the electrical part of an OLT line card. Then, the average energy saving of the OLT equipped with an electrical switch equals to:
$$1-\frac{l(t)\cdot p(e)}{L\cdot p(l)}$$
The efficiency of energy saving of this scheme depends on the ratio $p(e)/p(l)$.


\section{Performance evaluation}
In this section, we study the performance of the sleep control scheme of the OLT line cards for Poisson and non-Poison traffic. In the Markov chain model, we consider Poisson traffic arrival for the ease of analysis. The total number of line cards is assumed to be $4$ with the capacity of $10Gb/s$ for each line card. The time duration of each traffic scheduling cycle is considered to be $2ms$.
\begin{figure*}[ht!]
     \begin{center}
        \subfigure[Energy saving vs. arrival traffic rate]{%
            \label{fig:loadnum}
            \includegraphics[bb=60bp 20bp 950bp 830bp,clip,scale=0.18]{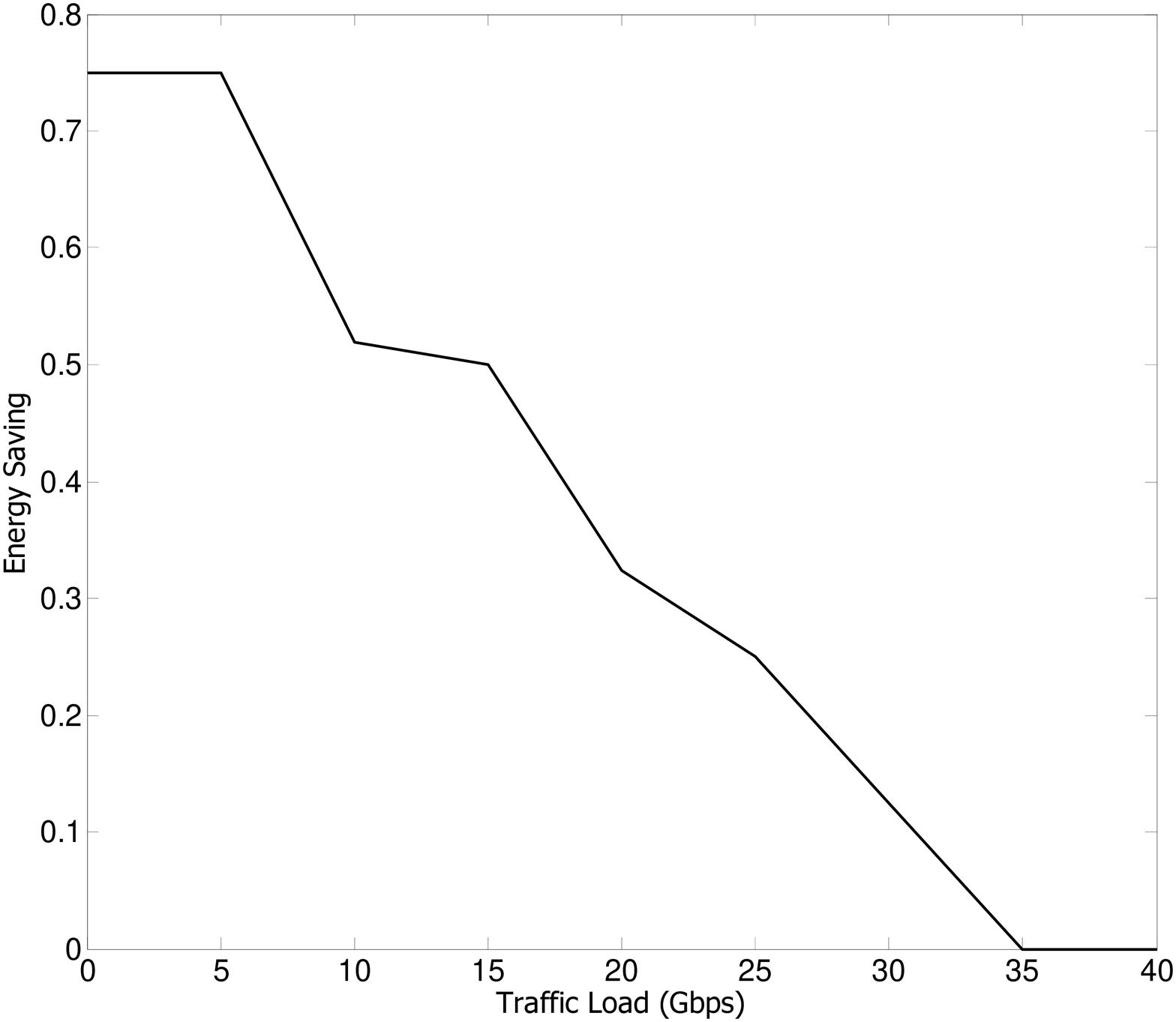}
        }%
        \centering
        \subfigure[Energy saving vs. $M$ listen cycles]{%
            \label{fig:Dlistennum}
            \includegraphics[bb=55bp 20bp 950bp 830bp,clip,scale=0.18]{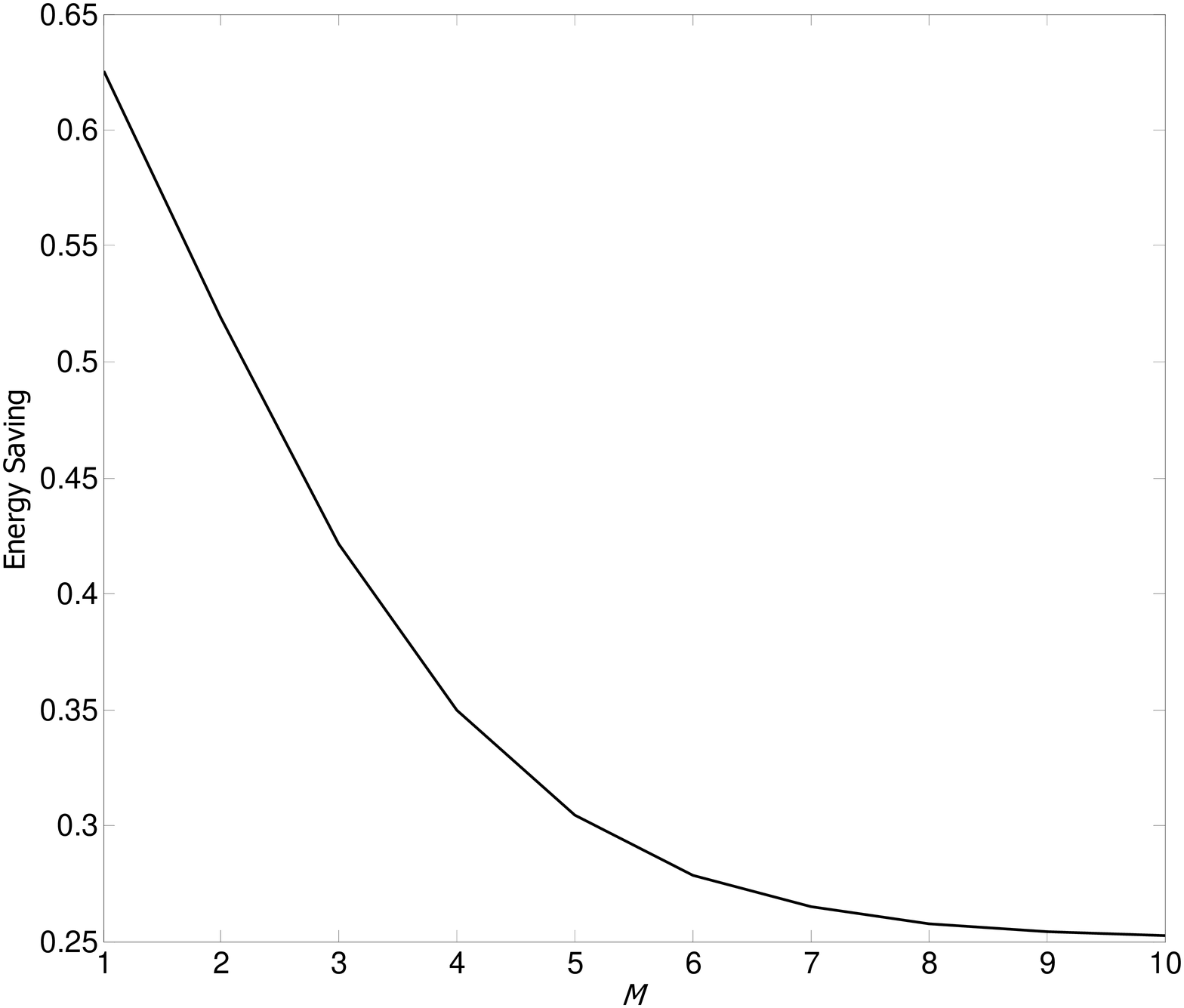}
        }%
        \subfigure[Energy saving vs. $N$ listen cycles]{%
           \label{fig:Ilistennum}
           \includegraphics[bb=60bp 20bp 950bp 830bp,clip,scale=0.18]{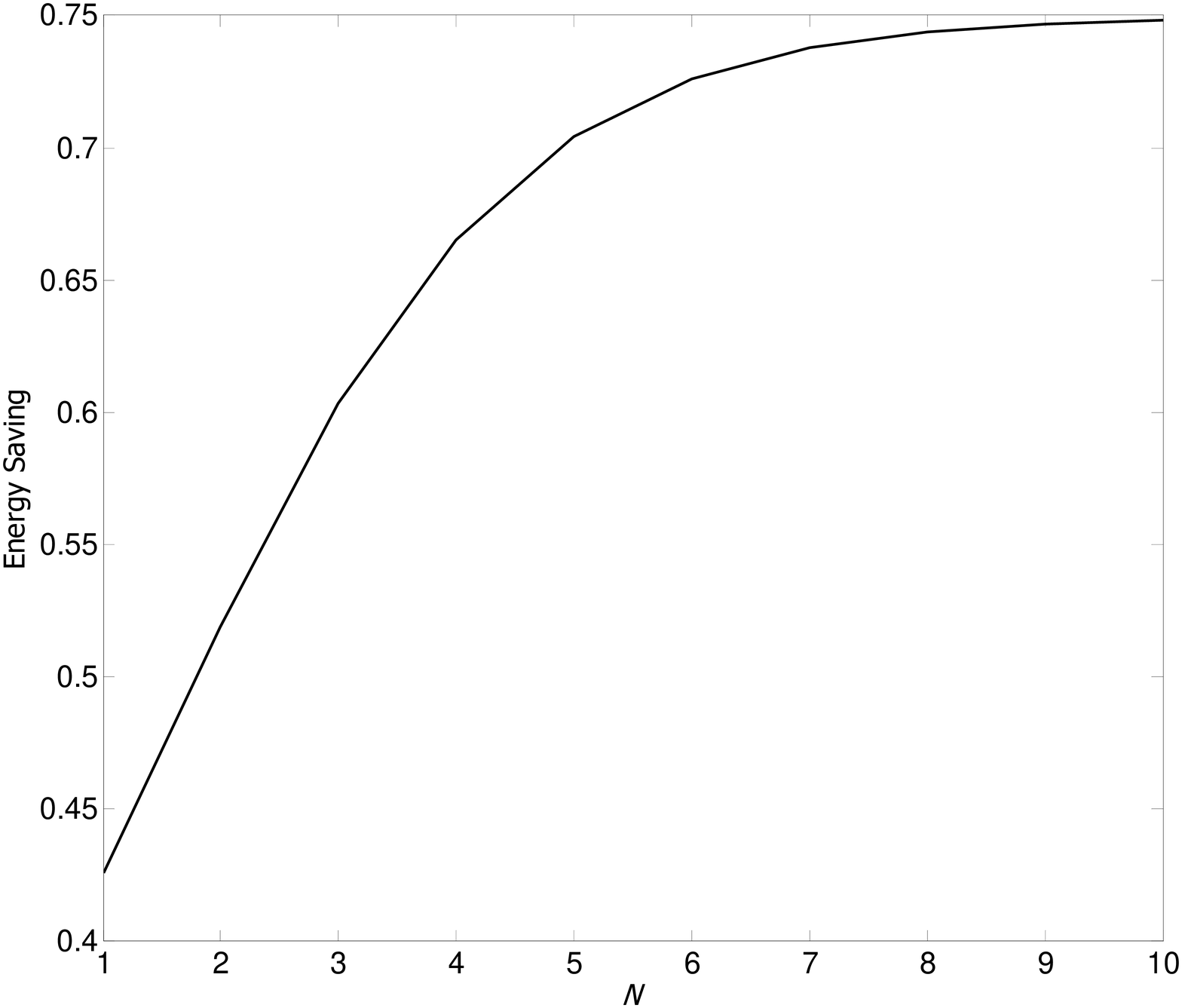}
        }%
\end{center}
    \caption{%
      Energy saving for Poisson traffic.
     }%
   \label{fig:numerical}
\end{figure*}

\begin{figure*}[ht!]
     \begin{center}
        \subfigure[Energy saving vs. arrival traffic rate]{%
            \label{fig:loadsim}
            \includegraphics[bb=60bp 20bp 950bp 830bp,clip,scale=0.18]{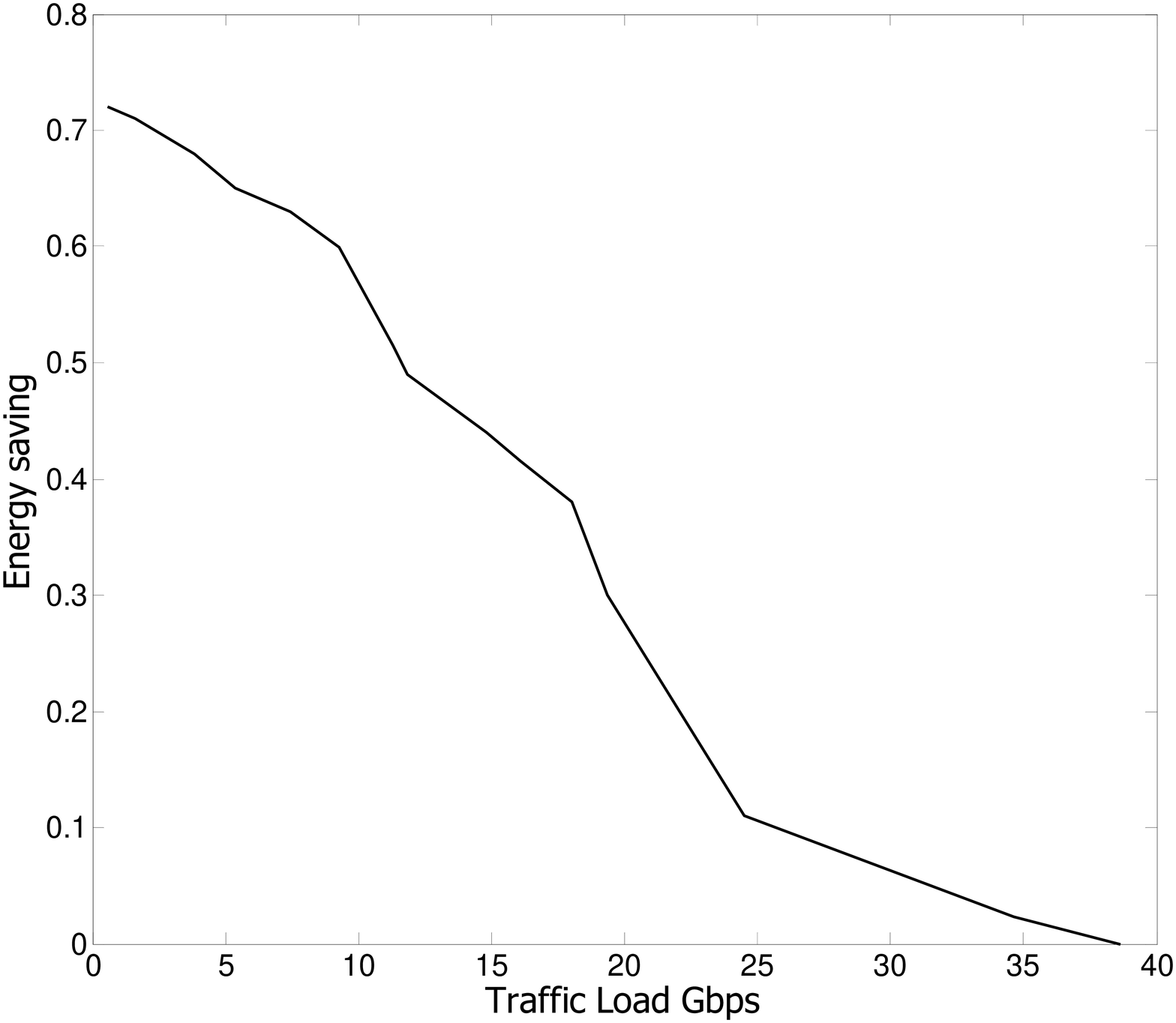}
        }%
        \subfigure[Energy saving vs. $M$ listen cycles]{%
            \label{fig:Dlistensim}
            \includegraphics[bb=55bp 20bp 950bp 830bp,clip,scale=0.18]{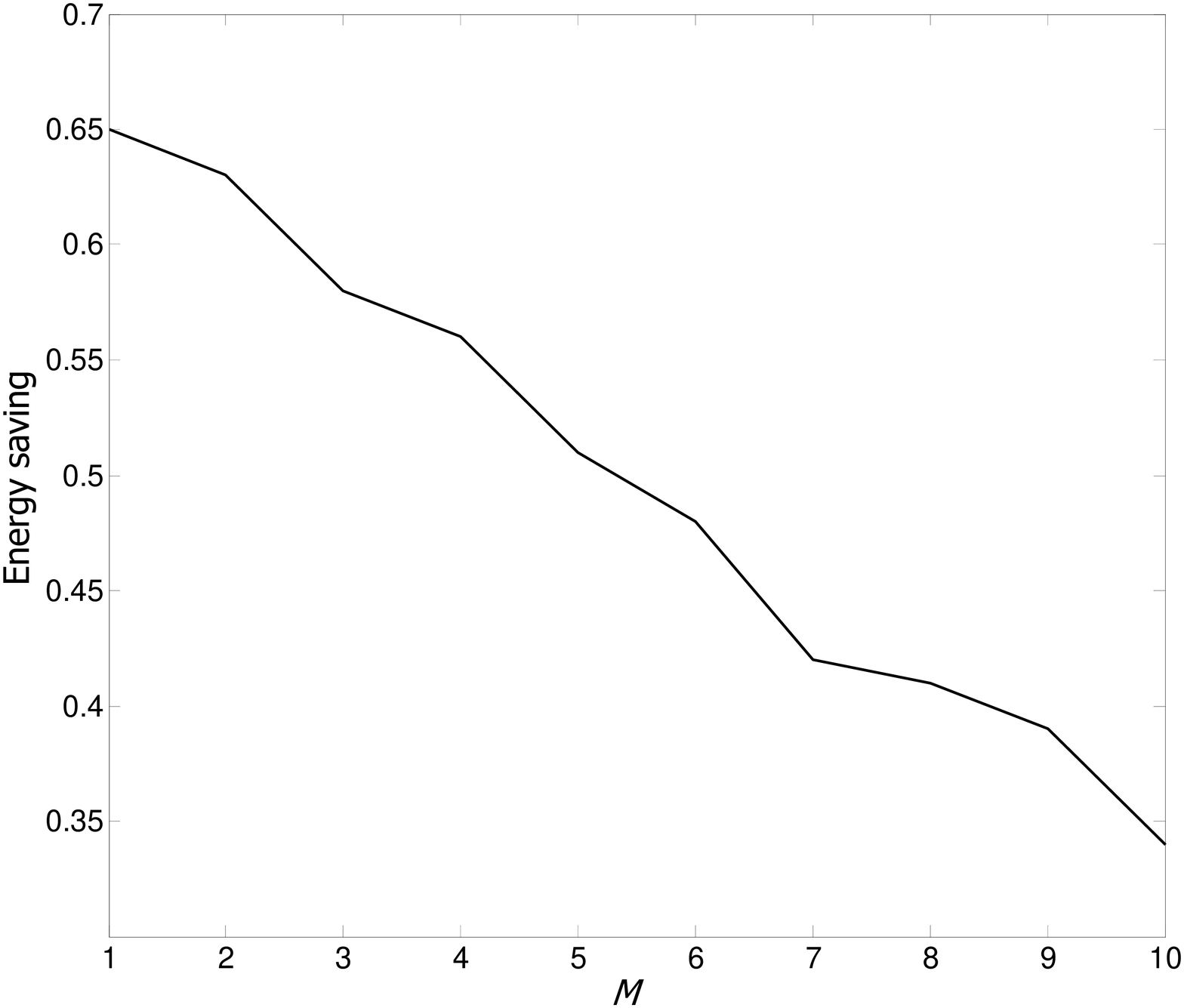}
        }%
        \subfigure[Energy saving and delay vs. $N$ listen cycles]{%
           \label{fig:Ilistensim}
           \includegraphics[bb=60bp 20bp 950bp 830bp,clip,scale=0.18]{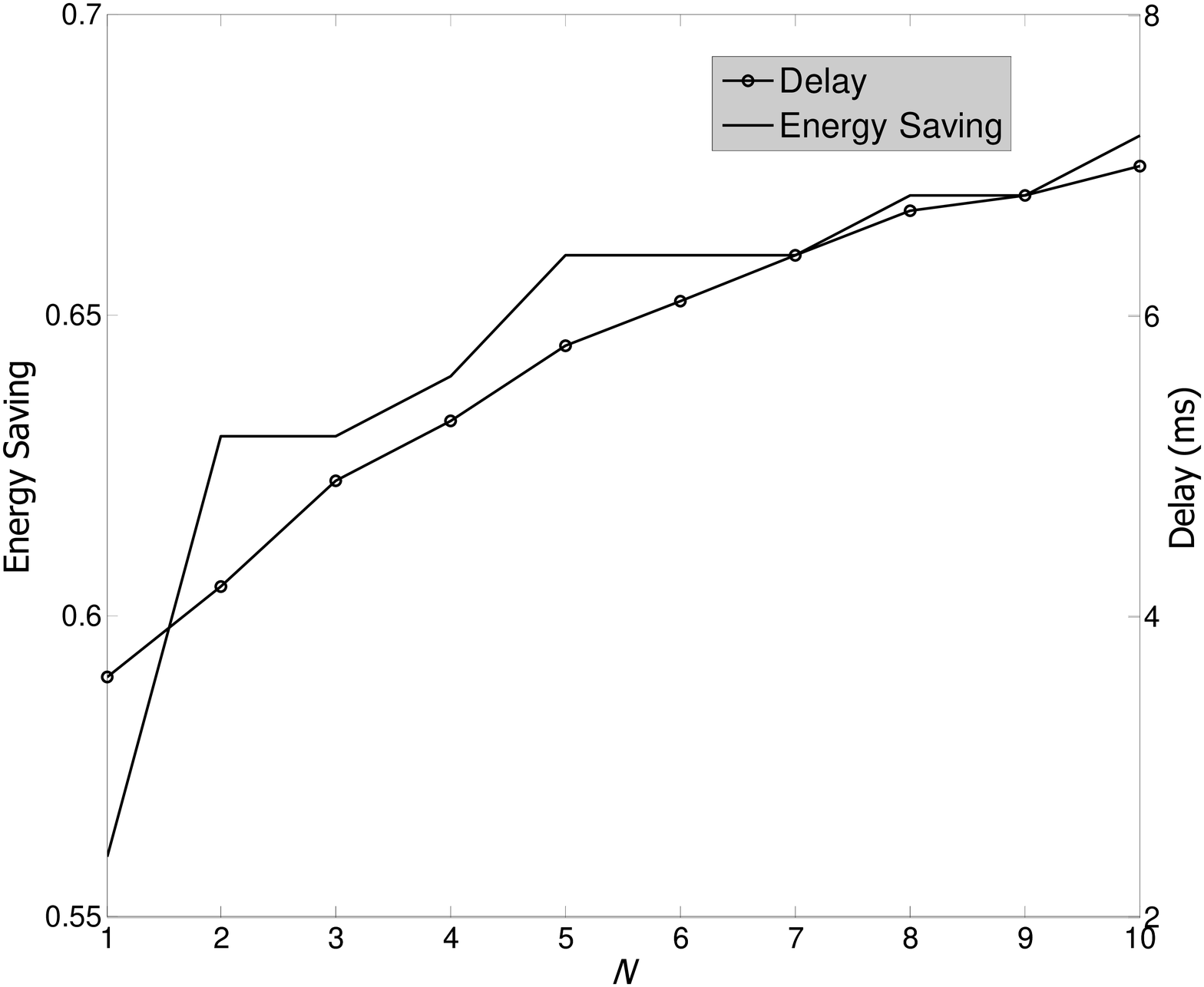}
        }%
\end{center}
    \caption{%
      Energy saving for self-similar traffic.
     }%
   \label{fig:sim}
\end{figure*}

Figure \ref{fig:numerical} depicts the numerical results of the Markov model. In Figure \ref{fig:loadnum}, the energy saving performance under different traffic arrival rates is illustrated. We assume that the maximum number of ``D'' listen cycles ($M$) as well as the maximum  number of ``I'' listen cycles ($N$) equal to $2$. The arrival traffic is increased up to $40Gb/s$. With the increase of traffic arrival, the energy saving decreases as the OLT chassis need to have more active line cards to support the  arrival traffic. When the arrival traffic is less than $10Gb/s$, one line card is sufficient to satisfy the traffic and the other $3$ line cards are shut down. Therefore, the maximum energy saving is achieved. 

Before decreasing the number of OLT line cards, the OLT chassis needs to monitor the traffic for the maximum of $M$ listen cycles. Figure \ref{fig:Dlistennum} shows the effect of changing the number of M on energy saving. With the increase of M, the OLT chassis needs to keep a larger number of line cards active for the time duration of $M \cdot T$. Thus, the energy saving decreases by increaing $M$.

$N$ is the maximum number of traffic cycles during which the OLT chassis needs to monitor the traffic before increasing the number of line cards. With the increase of $N$, the energy saving increases as the number of active line cards is smaller for the total duration of $N \cdot T$. 

We also study the performance of the proposed scheme for non-Poisson traffic. Since the actual network traffic with bursts is self-similar, we conduct our simulation for self-similar traffic with the Hurst parameter of $0.8$. The packet length is uniformly distributed between $64$ bytes and $1518$ bytes.

As it can be seen in Figure \ref{fig:loadsim}, the trend of energy saving performance by increasing the traffic load for self-similar traffic is similar to that of the Markov model for Poisson traffic. 

Figure \ref{fig:Dlistensim} shows the impact of different $M$ listen cycles on the system performance. Similar to the theoretical analysis of Poisson traffic, energy saving decreases as the number of listen cycles increases. However, increasing the $N$ cycles helps saving more energy. As mentioned earlier, during ``I'' listen cycles, the OLT chassis receives traffic load more than the available line cards. Before increasing the number of line cards, the OLT chassis needs to monitor the traffic for $N$ cycles. During theses cycles, the number of line cards stay the same, and the OLT buffers the extra traffic. The buffered packets encounter some delay as depicted in Figure \ref{fig:Ilistensim}. Therefore, proper setting of $N$ needs to be considered to save energy without impairing the QoS of the users.

\section{Conclusion}
We have proposed a novel energy-efficient OLT structure which adapts its power-on line cards to the real-time arrival traffic. Specifically, we have added a switch into the legacy OLT chassis to dynamically configure the connection between OLT line cards and ONUs. We first describe an OLT equipped with an $N\times N$ switch, and investigate the impacts of the power consumption of the optical switch and switch configuration time on the saved energy of the whole OLT chassis. Then, to avoid a dramatic cost increase, we advocate the use of opto-mechanical switches among all currently commercially-available switches, and further propose to use a cascaded $2\times 2$ switch structure. Our analysis demonstrates that the proposed OLT achieves significant power savings as compared to the legacy OLT.

\bibliographystyle{IEEETran}

\begin{thebibliography}{15}
\bibitem{BalEne09}
J.~Baliga, R.~Ayre, K.~Hinton, W.~Sorin, and R.~Tucker, ``{Energy consumption
  in optical IP networks},'' \emph{IEEE/OSA Journal of Lightwave Technology},
  vol.~27, no.~13, pp. 2391--2403, 2009.

\bibitem{pickavet2009worldwide}
M.~Pickavet,~\emph{et al.}, ``{Worldwide energy needs for ICT: the
  rise of power-aware networking},'' in \emph{2nd International Symposium on
  Advanced Networks and Telecommunication Systems}.\hskip 1em plus 0.5em minus
  0.4em\relax IEEE, 2008, pp. 1--3.

\bibitem{fettweisict}
G.~Fettweis and E.~Zimmermann, ``{ICT energy consumption-trends and
  challenges},'' in \emph{The 11th International Symposium on Wireless Personal
  Multimedia Communications}.vol.~2, no.~4, 2008.

\bibitem{BalEne08}
J.~Baliga, R.~Ayre, W.~Sorin, K.~Hinton, and R.~Tucker, ``{Energy consumption
  in access networks},'' in \emph{Optical Fiber Communication Conference and
  Exposition and The National Fiber Optic Engineers Conference}, 2008.

\bibitem{LanOnt08}
C.~Lange, M.~Braune, and N.~Gieschen, ``{On the energy consumption of FTTB and
  FTTH access networks},'' in \emph{National Fiber Optic Engineers Conference},
  2008.

\bibitem{ZhaNex09}
J.~Zhang, N.~Ansari, Y.~Luo, F.~Effenberger, and F.~Ye, ``Next-generation
  {PON}s: a performance investigation of candidate architectures for
  next-generation access stage 1,'' \emph{IEEE Communications Magazine},
  vol.~47, no.~8, pp. 49--57, August 2009.

\bibitem{ansari2013media}
N.~Ansari and J.~Zhang, \emph{Media Access Control and Resource Allocation for Next Generation Passive Optical Networks}, Springer, ISBN: ISBN: 978-1461439387, 2013.
  
%

\bibitem{ZhaTow11}
J.~Zhang and N.~Ansari, ``Towards energy-efficient {1G-EPON} and {10G-EPON}
  with sleep-aware {MAC} control and scheduling,'' \emph{IEEE
  Communications Magazine}, vol. 49, no. 2, pp. S33-S38, February 2011.



\bibitem{zhang2013standards}
J.~Zhang, M. Taheri, and N.~Ansari, ``Standards-compliant EPON sleep control for energy efficiency: Design and analysis,'' \emph{Journal of Optical Communications and Networking}, vol. 5, no. 7, pp. 677--685, 2013.

\bibitem{KubStu10}
R.~Kubo, J.~Kani, H.~Ujikawa, T.~Sakamoto, Y.~Fujimoto, N.~Yoshimoto, and
  H.~Hadama, ``{Study and demonstration of sleep and adaptive lLink rate control
  mechanisms for energy efficient 10G-EPON},'' \emph{IEEE/OSA Journal of
  Optical Communications and Networking}, vol.~2, no.~9, pp. 716--729, 2010.

\bibitem{WonSle09}
S.~Wong, L.~Valcarenghi, S.~Yen, D.~Campelo, S.~Yamashita, and L.~Kazovsky,
  ``{Sleep Mode for Energy Saving PONs: Advantages and Drawbacks},'' in
  \emph{2009 IEEE GLOBECOM Workshops}, 2009, pp. 1--6.

\bibitem{ChoEne10}
P.~Chowdhury, M.~Tornatore, and B.~Mukherjee, ``{On the energy efficiency of
  mixed-line-rate networks},'' in \emph{National Fiber Optic Engineers
  Conference}.\hskip 1em plus 0.5em minus 0.4em\relax IEEE, 2010, pp. 1--3.

\bibitem{taheri2014multi}
M.~Taheri and N.~Ansari, ``Multi-Power-Level Energy Saving Management for Passive Optical Networks,'' \emph{Journal of Optical Communications and Networking}, vol.~6, no.~11, pp.965--973, 2014.

\bibitem{Zhang2011}
J.~Zhang, T.~Wang, and N.~Ansari, ``Designing energy-efficient optical line terminal for TDM passive optical networks,'' in \emph{Sarnoff Symposium, 2011 34th IEEE}, pp.1--5, May 2011.

\bibitem{8023av}
{IEEE approved Draft Std P802.3av/D3.4}, 2009.

\bibitem{984}
{ITU-T Recommendation G.984 series}, http://www.itu.int/rec/T-REC-G/e.

\bibitem{opto}
{2x2 Opto-Mechanical Switch}, \\
{http://www.unitedoptronics.com/Datasheet/OpticalSwitch/OSW2x2.pdf}.
\bibitem{Vereecken2011}
W.~Vereecken,~\emph{et al.}, ``Power consumption in telecommunication networks: overview and reduction strategies,'' \emph{Communications Magazine, IEEE}, vol.~49, no.~6, pp.62--69, 2011.

\bibitem{MaxOpt03ofc}
X.~Ma,~\emph{et al.}, ``{Optical switching technology comparison: optical MEMS
  vs. other technologies},'' \emph{IEEE Commun. Mag}, vol.~41, no.~11, pp.
  16--23, 2003.
  
\end{thebibliography}

\end{document}